\documentclass[namedreferences]{SolarPhysics}
\usepackage[optionalrh,solaenum]{spr-sola-addons} 
\usepackage{graphicx}        
\usepackage{color}           
\usepackage{url}             

\newcommand{\etal}{{\it et al.}}

\renewcommand{\vec}[1]{ {\mathbf #1} }
\newcommand{\citext }[2]{\citeauthor{#1} (\citeyear{#1}, \citeyear{#2})}

\newcommand{\adv}{    {\it Adv. Space Res.}}

\newcommand{\aap}{    {\it Astron. Astrophys.}}

\newcommand{\apj}{    {\it Astrophys. J.}}

\newcommand{\solphys}{{\it Solar Phys.}}
\newcommand{\sovast}{ {\it Sov. Astron.}}

\begin{document}

\begin{article}

\begin{opening}

\title{Effect of collisions and magnetic convergence on electron acceleration and transport in reconnecting %
twisted solar flare loops}

\author{M.~\surname{Gordovskyy}$^{1}$\sep 
        P.K.~\surname{Browning}$^{1}$\sep 
		  E.P.~\surname{Kontar}$^{2}$\sep 
		  N.H.~\surname{Bian}$^{2}$
       }
\runningauthor{Gordovskyy et al.}
\runningtitle{Electron acceleration in twisted coronal loops}

   \institute{$^{1}$ Jodrell Bank Centre for Astrophysics, University of Manchester, Manchester M13 9PL, UK.
                     email: \url{mykola.gordovskyy@manchester.ac.uk} \\
				$^{2}$ School of Physics and Astronomy, University of Glasgow, Glasgow G12 8QQ, UK.
             }

\begin{abstract}
We study a model of particle acceleration coupled with 
an MHD model of magnetic reconnection in unstable twisted coronal loops. The kink instability 
leads to the formation of helical currents 
with strong parallel electric fields resulting in electron acceleration. The motion of 
electrons in the electric and magnetic fields of the reconnecting loop is investigated 
using a test-particle approach taking into account collisional scattering. 
We discuss the effects of Coulomb collisions and magnetic convergence near loop footpoints on the spatial distribution 
and energy spectra of high-energy electron populations and possible implications on the hard X-ray 
emission in solar flares. 
\end{abstract}


\end{opening}

\section{Introduction}

Most models of particle acceleration assume that particles gain energy in the corona and then are transported into 
the chromosphere where they produce non-thermal radiation. 
This assumption follows from the standard solar flare model (\textit{e.g.} \opencite{shib96}; \opencite{yosh98}), 
with the primary energy release site located above the 
top of a flaring loop or loop arcade.
However, the standard scenario faces a number of difficulties \cite{mkbr89,beho94,broe09}. 
The main problem is the large flux of electrons required to produce the observed hard X-ray (HXR) intensity. Also, 
unless precipitating electrons are accompanied by ions, they may result in a strong return current, creating additional 
electrodynamic challenges, such as preventing the electron beam from reaching the chromosphere.

It has been suggested recently, that re-acceleration of electrons in the chromosphere may help to overcome the ``number problem'' 
\cite{broe09}. In the proposed scenario electrons are initially accelerated in the corona and transported to the chromosphere. 
However, unlike in the standard thick target model (see \opencite{brow71}), electrons lose their energy much slower due to re-energization by the electric field in the chromosphere.
Indeed, this is possible even with comparably small field strength. In the presence of Coulomb collisions, the value of electric 
field $E$ required to accelerate electrons depends on their energy $\mathcal{E}_e$ as $E = q n/(e \mathcal{E}_e)$, 
where $q$ is an almost
constant parameter ($q \approx 0.017 \epsilon_0^{-2} e^3 k_B Z ln\Lambda$) (see \textit{e.g.} \opencite{plphbk}). 
Thus, for the typical 
chromospheric density of $10^{12}$ cm$^{-3}$, one needs a field of at least $30$ V m$^{-1}$ to accelerate electrons 
from $10$ eV (approximately thermal energies), but only $10^{-2}$ V m$^{-1}$ to accelerate electrons from $40$ keV. 
This electric field can be achieved even with the classical resistivity provided the current density is about 1 A m$^{-2}$.

Particle acceleration in fragmented electric fields distributed inside flaring loops can provide
another opportunity to solve the ``number problem'', yielding acceleration efficiencies broadly consistent with 
those deduced from HXR observations. Such an acceleration model would help to reduce energy lost during particle transport and to 
avoid some fundamental electrodynamic issues characteristic to acceleration models involving a single 
localized current layer, including the return current effect (\textit{e.g.} \opencite{knst77}) or ion-electron separation
(see \textit{e.g.} \opencite{zhgo04}). 

There are a few previous studies of distributed particle acceleration. \inlinecite{vlae04} considered acceleration of particles
by a system of fragmented unstable current sheets derived using the cellular automata model, with a particle
motion described by continuous-time random walks. Although the motion of individual particles is oversimplifield, this
model gives a good description of stochastic acceleration of particle population by a complex, fractal 
network of acceleration centres. \inlinecite{ture06} investigated particle 
acceleration using a stationary model of the stressed coronal magnetic field. 
\inlinecite{bibr08} used the kinetic approach to study stochastic acceleration by fluctuating fragmented 
electric fields in a turbulent reconnecting plasma and determined the relation between the distribution of current 
sheets and resulting particle energy spectra. 
More recently, Gordovskyy and Browning (2011, 2012)
considered proton and electron acceleration in twisted coronal loops using a time-dependent three-dimensional 
field configuration (see also \opencite{broe11}). These 
configurations are attractive as they are expected to be ubiquitous in the corona, as loops can be twisted 
by photospheric motions (see \textit{e.g.} \opencite{broe03}) or can emerge already twisted \cite{hooe11}. 
In any case, they contain excess magnetic 
energy which can be released if a loop becomes unstable.

Magnetic field convergence is expected near chromospheric footpoints of coronal loops and should have 
significant effect on particle motion.
In the present paper we examine the effects of both magnetic field convergence and Coulomb collisions 
on particle acceleration and re-acceleration in twisted loops, effects which have not been considered in
our previous works.

\section{Evolution of twisted magnetic loop and particle motion}

\begin{figure}    
\centerline{\includegraphics[width=1.0\textwidth,clip=]{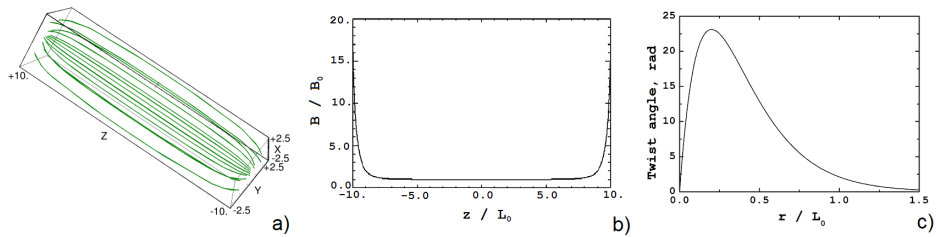}}
\caption{Initial magnetic field for the model with convergence near footpoint boundaries (panel a) and corresponding magnetic induction
on the $r=0$ axis (panel b). Panel (c) shows the final twist profile in the flux tube.}
\end{figure}

\begin{figure}    
\centerline{\includegraphics[width=0.8\textwidth,clip=]{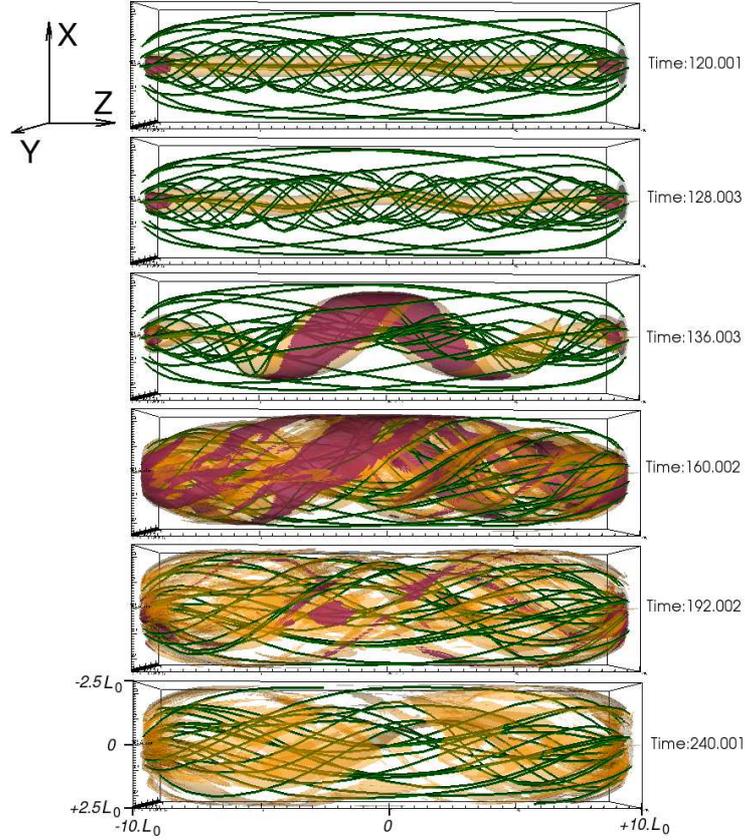}}
\caption{Evolution of magnetic field (green lines) and current density in the twisted flux tube with convergence 
near footpoints. Yellow and pink colours show
$|j|=4 j_0$ and $|j|=8 j_0$ iso-surfaces, respectively. Times shown are in the units of $t_0$.}
\end{figure}
\begin{figure}    
\centerline{\includegraphics[width=0.9\textwidth,clip=]{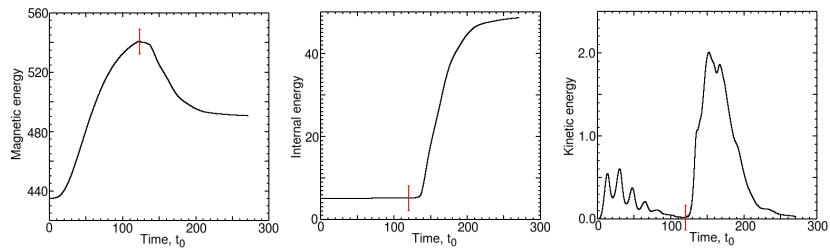}}
\caption{Variation of magnetic, kinetic and internal energies in twisted magnetic flux tube with the convergence 
near footpoint shown in Figure 2.
Verical dash at $t\approx 100 t_0$ denotes the time when footpoint rotation stops. Energy is measured in units of
$\frac 12 \mu_0^{-1} B_0^2 L_0^3$.}
\end{figure}

We consider twisted magnetic flux tube models with and without field convergence near footpoint boundaries. 

Magnetic reconnection in initially cylindrical flux tubes has been extensively studied using
analytical initial equilibria \cite{broe08,hooe09}. These models have been already employed to study particle acceleration in 
\citext{gobr11}{gobr12}, so that we do not discuss them further here.

A configuration with twisted magnetic field converging near footpoints can not be set up analytically. 
Therefore, it is derived numerically, using ideal MHD, by twisting the footpoints of an initially current-free magnetic 
flux tube. We use the simulation domain $x=[-5 L_0;5 L_0]$, $y=[-5 L_0;5 L_0]$, $z=[-10 L_0;10 L_0]$. In this model 
scaling parameters are the same as in \inlinecite{gobr12}: the scale length is $L_0 = 10^6$ m, the characteristic magnetic 
induction is $B_0 = 4\times 10^{-3}$ T and the characteristic density is $\rho_0 = 3.3\times 10^{-12}$ kg m$^{-3}$; 
which yields the characteristic Alfv\'en velocity $v_0 = 1.95\times 10^6$ m s$^{-1}$ and the characteristic 
timescale $t_0 = 0.53$ s. 

Initially, the field is potential, formed by two point "sources" located at $\vec{r}_1=\{0;0;-10.5 L_0\}$ 
and $\vec{r}_2=\{0;0;10.5 L_0\}$ outside the simulation domain (Figure 1a):
\begin{equation}
\vec{B}(\vec{r}) = B_s\left[\frac{\vec{r}-\vec{r}_1}{|\vec{r}-\vec{r}_1|^3}-\frac{\vec{r}-\vec{r}_2}{|\vec{r}-\vec{r}_2|^3}\right].
\end{equation}
This yields a flux tube with the magnetic field at footpoints nearly 10 times higher than the field at the centre (Figure 1b).
The initial density and pressure in the MHD simulations are uniform: $\rho(t=0) = \rho_0$ and 
$p(t=0) = 5\times 10^{-3} \mu_0^{-1} B_0^2$, \textit{i.e.} it is magnetically dominated plasma. 

The twist is introduced using the azimuthal velocity at the $z=\pm10L_0$ boundaries (corresponding to the
photospheric footpoints). The angular velocity depends on the radius as $\omega \sim r^2/r_0^2 \exp(-r/r_0)$, which
is generally consistent with observed sunspot rotation (\textit{e.g.} \opencite{gopa77}), where $r_0 = 0.2 L_0$. The corresponding 
twist (defined as $ \phi(r,t)=\int_0^t \omega(r,t)dt$) is shown in Figure 2c. 
The resulting configuration (at $t=100t_0$) has a maximum twist of approximately $7\pi$ and is nearly force-free but 
is, apparently, unstable: the kink instability develops at $t \approx 120 t_0$. 

It should be noted, that this time
is much shorter than realistic timescale of development of kink instability. Indeed, twisted coronal loops can remain in 
equilibrium for up to $10^3 - 10^4$ Alfv\'en times. However, since we are interested in the magnetic reconnection occuring after 
the loop kinks, the development of the instability is speeded up by the velocity noise (see \opencite{gobr11}).

Further evolution of the twisted flux tube once it becomes unstable is studied using resistive MHD simulations 
with the Lare3D code \cite{arbe01}
(see also \citext{gobr11}{gobr12} for more details).
The resistivity in this part of the simulation is non-uniform and depends on local current density as
\begin{equation}
\eta(j) = \left\{ 
\begin{array}{ll}
0 & j < j_{cr}\\
\eta_1 & j\geq j_{cr}
\end{array}
\right.
\end{equation}
The critical current in the present simulations is  $j_{cr}=8 \mu_0^{-1} B_0 L_0^{-1}$ and the resistivity
value is $\eta_1 = 10^{-3} L_0^2 t_0^{-1}$ (which is, essentialy, the inverse Lundquist number). 

Magnetic field and current density evolution for
the model with magnetic convergence near footpoints is shown in Figure 2, while Figure 3 shows the variation of 
magnetic energy $E_M$, kinetic energy $E_K$ and internal energy $E_I$ defined by following integrals over the 
domain volume: 

\begin{eqnarray}
E_M(t) &=& \frac 1{2\mu_0}\int\limits_{\rm volume} B^2 d\Omega \nonumber \\
E_K(t) &=& \frac 1{2}\int\limits_{\rm volume} \rho v^2 d\Omega \nonumber \\
E_I(t) &=& \frac 1{\gamma-1}\int\limits_{\rm volume} p d\Omega, \nonumber 
\end{eqnarray}
where $d\Omega$ is the elementary volume.
The field and current for initially
cylindrical flux tube can be seen in Gordovskyy and Browning (2011, Figures 3 and 4). It can be seen 
that both configurations (with and
without magnetic convergence) demonstrate similar behaviour. In both cases, when the twisted flux tube kinks, the
current density rapidly increases forming a helically-shaped structure. The decrease in magnetic energy at this stage is 
due to the dissipation of the azymuthal component of magnetic field and reconnection between twisted loop's 
field lines and ambient field lines. Gradually, the current distribution becomes more filamentary and the 
current density decreases. Reconnection stops after about $100 t_0$ after kink instability occurs.
During the rapid reconnection ($t=130\; -\; 180\;t_0$, see Figure  3), the electric field in the 
system is order of $10^2$ V m$^{-1}$, although it
occupies only a tiny fraction of the model volume (see Gordovskyy and Browning, 2012). The main difference between the 
initially cylindrical configuration and one with converging field is that, in the latter case, stronger magnetic field 
results in substantially enhanced current density near the footpoints.  

\begin{figure}    
\centerline{\includegraphics[width=0.5\textwidth,clip=]{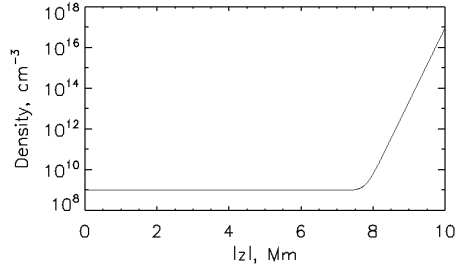}}
\caption{Density profile within the flux tube used to account for the effect of Coloumb collisions. The ``coronal'' part 
($|z| < 7.5$ Mm) of the flux tube has a constant density of $10^9$ cm$^{-3}$ while the ``chromospheric'' part 
($|z| = 8\; - \; 10$ Mm) has an exponential denisty distribution with the scale height of about $170$ km.}
\end{figure}

The obtained time-dependent electric and magnetic field configurations are used to calculate test-particle trajectories. 
Initially, protons and electrons in the model are uniformly distributed in the domain volume, with Maxwellian velocity 
distribution corresponding to $T=0.8$ MK and an isotropic pitch-angle distribution.
Since the scale-length of our MHD model is much larger than particle gyroradii, we use the
relativistic guiding centre approximation to calculate particle trajectories (see \opencite{nort63}):

\begin{eqnarray}
\frac {d \bf r}{dt} &=& {\bf u} + \frac{\gamma (v_{||})}\gamma {\bf b}\\
{\bf u} &=& {\bf u}_E + %
\frac mq \frac {(\gamma v_{||})^2}{\gamma \kappa^2 B} [{\bf b} \times ({\bf b}\cdot{\bf \nabla}){\bf b}] + %
\frac mq \frac \mu {\gamma \kappa^2 B} [{\bf b} \times ({\bf \nabla} (\kappa B))]\\
\frac{d (\gamma v_{||})}{dt} &=& \frac qm {\bf E}\cdot {\bf b} - \frac \mu \gamma ({\bf b} \cdot {\bf \nabla}(\kappa B))%
+ \left[ \mathcal{V} \left(\frac{\delta \alpha}{\delta t}\right)_{pa} %
- \frac {(\gamma v_{||})}{\mathcal{V}} \frac{an}{\mathcal{V}^2}\right]_{coll}\\
\frac{d\mu }{dt} &=& \left[(\gamma v_{||})\sqrt{\frac{2 \mu}{B(1-(\gamma v_{||})^2/\mathcal{V}^2)}}%
- \frac {2\mu}{\mathcal{V}} \frac{an}{\mathcal{V}^2} \left(\frac{\delta \alpha}{\delta t}\right)_c \right]_{coll}\\
\gamma &=& \sqrt{\frac{c^2 +(\gamma v_{||})^2 + 2 \mu B}{c^2 - u_E^2}},
\end{eqnarray}
where $v_{||}$ is the particle velocity along the magnetic field ($\gamma v_{||}$ is treated as a single variable 
for the sake of convenience), and ${\bf u}$ is the perpendicular velocity. The drift velocity ${\bf u}_E$ is 
defined as ${\bf u}_E=\frac {{\bf E}\times{\bf B}}{B^2}$. Also, ${\bf b}$ is the 
magnetic field direction vector ${\bf b} = {\bf B}/B$, $\mu$ is the magnetic moment defined as
$\mu = \frac 12 (\gamma v_g)^2 / B$, $\gamma$ is the Lorenz factor ($\gamma = (1-v^2/c^2)^{-1/2}$); 
$\kappa = \sqrt{1-u^2_E/c^2}$. 

The terms in $[\;\;]_{coll}$ brackets are to account for collisional energy losses and pitch-angle scattering,
which are not usually incorporated in test-particle models.
For typical coronal and chromospheric densities collisions are predominantly important for particles with 
enegies in $1-100$ keV range and, therefore, relativistic effects in the collisional terms can be 
ignored. 

The experiments including Coulomb collisions are performed 
with the density profile shown in Figure 4, where the region with increasing density ($|z| = 8...10$ Mm) represents the chromosphere.
Collisional energy losses are taken into account through the decrease of the particle 
velocity $\mathcal{V}$ as follows:
\begin{equation}
\frac{d\mathcal{V} }{dt }= -a\frac n {\mathcal{V}^2}, 
\end{equation}
where the particle velocity for the purpose of Coulomb collisions is defined as 
$\mathcal{V}=\sqrt{v_{||}^2+u_{\rm g}^2}$. (The drift velocity $\vec{u}$ is defined 
solely by local field parameters and represents predominantly bulk plasma flow, 
and, therefore, is excluded from the collisional terms.) 
Here $n$ is the ambient plasma density and $a$ is the constant $a=2q/m_e^2$; for typical chromospheric 
parameters $q=2\pi e^4\;\ln \Lambda$ is approximately $2.87\times 10^{-12}$~eV$^2$~cm$^2$ \cite{sysh72}. 
Essentially, this parameter determines how deep an accelerated particle can penetrate into dense plasma; in case of 
an uniform density the stopping depth can be evaluated as follows:
\[
\left(\frac{S_{\max}}{1Mm}\right) \approx 1.75\left(\frac{E_{initial}}{1keV}\right)^2 \left(\frac{n}{10^9 cm^{-3}}\right)^{-1}.
\]

Pitch-angle scattering is implemented through random changes of test-particle pitch-angle ratio 
$\alpha = \cos \theta = v_{||} / \mathcal{V}$ as
\begin{equation}
\left(\frac{d\alpha}{dt}\right)_c = \Pi \left(dt \frac {an}{\mathcal{V}^3} \frac{1-(\alpha+\Delta \alpha/2)^2}%
{\Delta \alpha^2}\right)%
 \Delta \alpha - \Pi \left(dt \frac {an}{\mathcal{V}^3} \frac{1-(\alpha-\Delta \alpha/2)^2}{\Delta \alpha^2}\right) %
\Delta \alpha,
\end{equation}
where $\Pi(x)$ is the operator yielding $1$ or $0$ with the probability $x$ or $1-x$, respectively.
Provided $\Delta \alpha$ is sufficiently small, 
the pitch-angle distribution evolves as 
\[
\frac {\partial N(E,\alpha)}{ \partial t} = an \mathcal{V}^{-3} \frac {\partial }{ \partial \alpha}%
\left( (1-\alpha^2) \frac {\partial N(E,\alpha)}{\partial \alpha }\right),
\]
as expected.

In all the considered test-particle simulations we use open boundaries (\textit{i.e.} particles are free to leave the domain; 
their trajectories are calculated until they cross a boundary). 

\section{Non-thermal electrons in a twisted coronal loop}

Here we mostly focus on the spatial distribution of non-thermal electrons, 
which have direct implications for observations.
The energy spectra of particles accelerated in initially cylindrical twisted magnetic flux tubes have been discussed by
\inlinecite{gobr11} and \inlinecite{broe11}. In the present paper we compare those with the energy spectra of electrons 
accelerated in the presence of magnetic convergence and Coulomb collisions.

\begin{figure}    
\centerline{\includegraphics[width=1.0\textwidth,clip=]{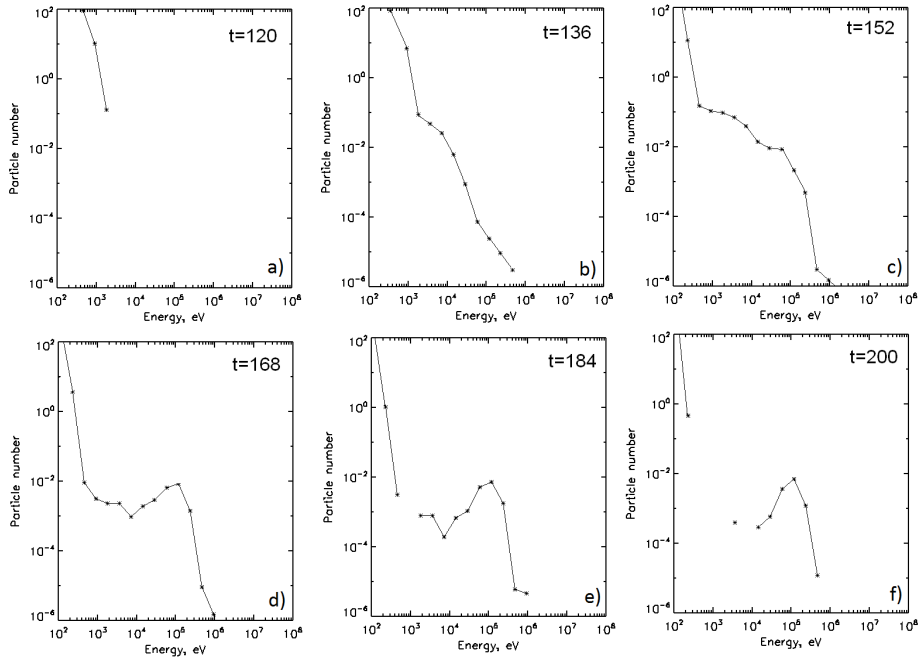}}
\caption{Energy spectra of electrons in the model with magnetic convergence and Coulomb collisions. 
Times (normalized by $t_0$) are shown in the panels. The shape of the distribution below 1 keV is not realistic 
on panels (b-f) (see text for details).}
\end{figure}

The electron energy distributions (Figure 5) at the early stages of reconnection 
are rather similar to those obtained in simulations with 
no magnetic convergence and no collisions \cite{gobr11}. Thus, during the fast energy release stage 
($t=130-150 t_0$) the spectra are combinations of a Maxwellian thermal distribution and nearly power-law
high-energy tail. However, at the later stages ($t=150-250 t_0$), when electric fields are gradually 
decaying and the collisions become dominant, the part of the spectra around few keV becomes harder and, at some point,
a gap appears between the thermal part and high-energy part of the spectra. This is similar to the energy
spectra of electrons precipitating in the thick-target model.

It should be noted, that the Equation (8) is valid only for non-thermal electrons (\textit{i.e.} when $\mathcal{E}_e \gg kT$ and,
therefore, the shapes of spectra in Figure 5(b-f) below $1$~keV are not realistic, although the total number of 
``thermal'' electrons is correct. This should not affect the behaviour of particles at higher energies.

Now let us consider spatial distribution of energetic electrons.
The distributions of electrons with energies $>5$ keV along the flaring loop are shown in the Figure 6 for three 
different cases. They are calculated assuming the electron populations are symmetric in respect of the loop 
mid-plane $z=0$. All three distributions 
correspond to the stage of fast energy release in the MHD model. It can be seen that in the 
case of initially cylindrical flux tube, high-energy particles are nearly uniformly distributed along the reconnecting 
loop, which can be interpreted as the result of rather uniform distribution of strong electric field ``islands''
along the loop's length. The total number of particles in this case is quite low because most electrons quickly get to 
one of the boundaries and leave the domain. 

In the case with converging magnetic field near footpoint boundaries high-energy electrons are rather uniformly
distributed around the centre of the flux tube, but with some deficit close to the footpoints. This deficit is quite
surprising, as one would expect a higher number of particles in these regions: particles are slowing down in stronger 
field and would spend more time near these boundaries. Apparently, the deficit is caused by more efficient acceleration
due to stronger electric fields in these regions (see Section 2).

In the case with magnetic convergence and Coulomb collisions the majority of the electrons are 
located near the footpoints, which indicates that test-particles spend most of their time moving through 
the dense plasma representing the chromosphere. This phenomenon might be explained as the result of simultaneous
action of three different effects: the converging magnetic field leads to high pitch-angles (\textit{i.e.} $u_g > v_{||}$), which, 
along with collisional deceleration, prevents majority of electrons from escaping into central region of the flux tube
or outside. At the same time, electrons do not thermalize completely as they are accelerated by direct electric field.

\begin{figure}    
\centerline{\includegraphics[width=1.0\textwidth,clip=]{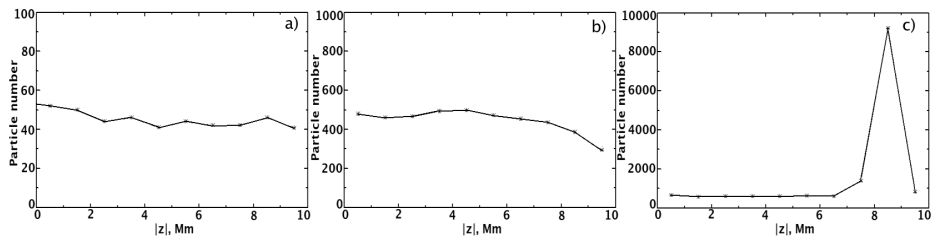}}
\caption{Distribution of electrons with energies $\mathcal{E} > 5$ keV along the flux tube. Here $z=0$ corresponds to the
centre of the flux tube and $|z|=10\, L_0$ corresponds to the footpoints.  
Panel (a) is for the initially cylindrical flux tube,
panel (b) is for the case with convergence near footpoints, panel (c) is for the case with magnetic convergence and Coulomb collisions. All the distributions correspond to the stage of rapid energy release in the MHD simulations.}
\end{figure}

\begin{figure}    
\centerline{\includegraphics[width=1.0\textwidth,clip=]{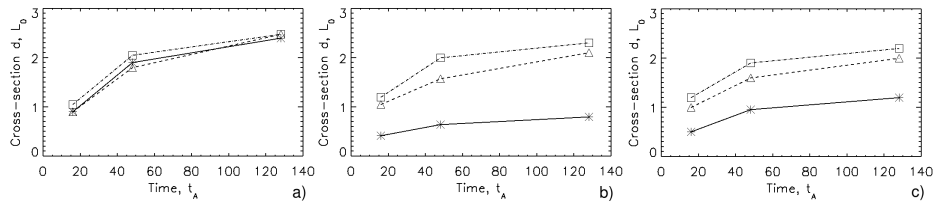}}
\caption{Variation of cross-section of the volume occupied by high-energy electrons. Panel (a) is for the 
initially cylindrical flux tube, panel (b) is for the case with convergence near footpoints, panel (c) is 
for the case with magnetic convergence and Coulomb collisions. Solid lines are for the region 
$|z|=8L_0...10L_0$, dashed lines are for $|z|=4L_0...6L_0$, dot-dashed lines are for central region 
$|z| < 2L_0$. Horizontal axis shows the time after onset of reconnection. (It should be noted that 
speed of reconnection in the models in the case (a) and in the case (b-c) can be slightly different and, 
therefore, times shown in panel (a) may correspond to slightly different phase compared to panels (b) and (c).)
The shape of cross-section is determined visually (as it has quite sharp edges, see Gordovskyy and Browning, 2011), 
the size is defined as average diameter from two measurements, in x- and y- directions.}
\end{figure}

In our previous paper \cite{gobr12} we show that the cross-section of the volume occupied by 
accelerated particles increases with time. This happens due to the reconnection between magnetic field of twisted loop
and the ambient magnetic field. Let us now examine how the cross-section sizes change with time at different distances 
from loop footpoints , in the three different models (Figure 7).

It can be seen that in the case of an initially cylindrical flux tube the centre and the footpoints regions 
expand in the same way. Cross-section diameters increase from $\approx 1 L_0$ to $\approx 2.5 L_0$
within $120 t_0$. 
In the case with field convergence, the central region expands nearly at the same rate
as the footpoint regions. Thus, around $z=0$ the cross-section of the volume occupied by high-energy
particles increases from $\approx 1.2 L_0$ to $\approx 2.3 L_0$ within about $100 t_0$, while the region 
between $|z| = 8...10 L_0$ expands from $0.4 L_0$ to $0.7 L_0$. 

With Coulomb collisions taken into account,
the footpoint sizes expand faster: the cross-section of the volume between $|z| = 8...10 L_0$
increases from $0.5 L_0$ to $\approx 1.2 L_0$ compared with the increase from $1.2 L_0$ to $2.2 L_0$ near
the centre of the flux tube. One of the possible explanations for this effect is strong pitch-angle scattering leading
to a wider spread of particles in radial direction.

The radial expansion of the volume occupied by particles should result in the increase of the radial size of non-thermal 
hard X-ray sources observed in solar flares. Recent observations by \inlinecite{kone10} and \inlinecite{kone11} show that
this expansion may be happening in solar flares and the observed pace of expansion is comparable to what is seen in the
numerical experiments.

\section{Summary}

In this paper we present preliminary results concerning electron acceleration in a kink unstable twisted coronal loop. 
For this purpose a model of twisted magnetic flux tube with the field convergence near its footpoints was developed 
using ideal MHD simulations. Further development of the kink instability resulting in magnetic reconnection and fast energy 
release was investigate using resistive MHD simulations with non-uniform resistivity.
We formulated a set of relativistic gyro-kinetic equations taking into account energy losses and pitch-angle scattering due to 
Coulomb collisions and, based on the developed MHD model, studied electron acceleration during magnetic reconnection in twisted 
magnetic flux tube.
  
The considered reconnection scenario appears to be an effective particle 
accelerator. Taking into account scaling parameters 
($\mathcal{E}_0 = 1/2 \, B_0^2/\mu_0 \, L_0^3 \approx 6\times 10^{18}$ J, 
$t_0 \approx 0.53$s) once can estimate the total energy released during the reconnection. In the case of twisted 
magnetic flux tube with convergence near footpoints there is $\approx 3\times 10^{20}$ J released during $\approx 50$ s.
The efficiency of particle acceleration, which strongly depends on the resistivity in these models, is quite low:
approximately $5-7\%$ of the total energy released is carried by accelerated particles.  
This, along with the fact that the magnetic reconnection occurs in rather simple single-loop configuration, makes this model 
comparable to small self-contained flares \cite{asce09}. We emphasize, however, that all these estimations are subject to
the scaling parameters.

The considered electron acceleration model is quite attractive because the
particle acceleration is not localized at the top of the loop but occurs along the whole loop length,
including the chromosphere.

The considered model yields number of observational implications. The main feature is expansion of the column
occupied by high-energy particles in radial direction. Taking into account recent progress in
determination of geometric properties of hard X-ray sources \cite{kone10,bako11}, this feature can be used as an observational 
test for this sort of models. However, more simulations are needed to investigate this reconnection and
acceleration scenario with wider range of parameters.

\begin{acks}
This work is supported by the Science and Technology Facilities Council (UK). Computational facilities have
been provided by the UK MHD Consortium. Financial support by the European Commission through the HESPE Network
is gratefully acknowledged.
\end{acks}

\end{article} 


\begin{thebibliography}{00}
\bibitem[\protect\citeauthoryear{Arber \etal}{2001}]{arbe01} Arber, T.D., Longbottom, A.W., Gerrard, C.L., %
Milne, A.M.: 2001, {\it J.Comput.Phys.} {\bf  171}, 151.
\bibitem[\protect\citeauthoryear{Aschwanden \etal}{2009}]{asce09} Aschwanden, M.J., Wuelser, J.P., Nitta, N.V., %
Lemen, J.R.: 2009, {\solphys } {\bf 256}, 3. doi: 10.1007/s11207-009-9347-4
\bibitem[\protect\citeauthoryear{Battaglia and Kontar}{2011}]{bako11} Battaglia, M., Kontar, E.P.: 2011, {\apj } {\bf 735}, 42.
\bibitem[\protect\citeauthoryear{Benka and Holman}{1994}]{beho94} Benka, S.G., Holman, G.D.: 1994, {\apj } {\bf 435}, 469.
\bibitem[\protect\citeauthoryear{Bian and Browning}{2008}]{bibr08} Bian, N.H., Browning, P.K.: 2008, {\apj } {\bf 687}, L111.
\bibitem[\protect\citeauthoryear{Brown \etal}{2003}]{broe03} Brown, D.S., Nightingale, R.W., Alexander, D., %
Schrijver, C.J., Metcalf, T.R., Shine, R.A., Title, A.M., Wolfson, C.J.: 2003, {\solphys } {\bf 216}, 79. %
 doi: 10.1023/A:1026138413791
\bibitem[\protect\citeauthoryear{Brown}{1971}]{brow71} Brown, J.C.: 1971, {\solphys } {\bf 18}, 489.%
 doi: 10.1007/BF00149070
\bibitem[\protect\citeauthoryear{Brown \etal}{2009}]{broe09} Brown, J.C., Turkmani, R., Kontar, E.P., MacKinnon, A.L., %
Vlahos, L.: 2009, {\aap } {\bf 508}, 993.
\bibitem[\protect\citeauthoryear{Browning \etal}{2008}]{broe08} Browning, P.K., Gerrard, C., Hood, A.W., %
Kevis, R., van der Linden, R.A.M.: 2008, {\aap } {\bf 485}, 837.
\bibitem[\protect\citeauthoryear{Browning \etal}{2011}]{broe11} Browning, P.K., Gordovskyy, M., Stanier, A., Hood, A.W., %
Dalla, S.: 2011, {\it Plasma Phys. Contr. Fusion}, {\bf 53}, 124030.
\bibitem[\protect\citeauthoryear{Gopasyuk}{1977}]{gopa77} Gopasyuk, S.I.: 1977, {\it Izv. Crimean Astr. Obs.}, {\bf 57}, 107.
\bibitem[\protect\citeauthoryear{Gordovskyy and Browning}{2011}]{gobr11} Gordovskyy, M., Browning, P.K.: 2011, %
{\apj } {\bf 729}, 101.
\bibitem[\protect\citeauthoryear{Gordovskyy and Browning}{2012}]{gobr12} Gordovskyy, M., Browning, P.K.: 2012, %
{\solphys }, {\bf 277}, 299. doi: 10.1007/s11207-011-9900-9
\bibitem[\protect\citeauthoryear{Hood, Browning, and van der Linden}{2009}]{hooe09} Hood, A.W., Browning, P.K., %
van der Linden, R.A.M.: 2009, {\aap } {\bf 506}, 913.
\bibitem[\protect\citeauthoryear{Hood, Archontis, and MacTaggart}{2012}]{hooe11} Hood, A.W, Archontis, V., MacTaggart, D.: %
2012, {\solphys }, {\bf 278}, 3. doi: 10.1007/s11207-011-9745-2 
\bibitem[\protect\citeauthoryear{Knight and Sturrock}{1977}]{knst77} Knight, J.W., Sturrock, P.A.: %
1977, {\apj }, {\bf 218}, 306. 
\bibitem[\protect\citeauthoryear{Kontar \etal}{2010}]{kone10} Kontar, E.P., Hannah, I.G., Jeffrey, N.L.S., %
Battaglia, M.: 2010, {\apj } {\bf 717}, 250.
\bibitem[\protect\citeauthoryear{Kontar, Hannah, and Bian}{2011}]{kone11} Kontar, E.P., Hannah, I.G., %
Bian, N.H.: 2011, {\apj } {\bf 730}, L22.
\bibitem[\protect\citeauthoryear{MacKinnon and Brown}{1989}]{mkbr89} MacKinnon, A.L., Brown, J.C.: 1989, {\solphys } {\bf 122}, 303.%
 doi: 10.1007/BF00912997
\bibitem[\protect\citeauthoryear{Northrop}{1963}]{nort63} Northrop, T.: 1963, %
{\it The Adiabatic Motion of Charged Particles}, Interscience, New York, p.32.
\bibitem[\protect\citeauthoryear{Shibata}{1996}]{shib96} Shibata, K.: 1996, {\adv } {\bf 17}, 9.
\bibitem[\protect\citeauthoryear{Spitzer}{1962}]{plphbk} Spitzer, L.: 1962, %
{\it Physics of fully ionized gases}, Interscience, New York.
\bibitem[\protect\citeauthoryear{Syrovatskii and Shmeleva}{1972}]{sysh72} Syrovatskii, S.I., Shmeleva O.P.: 1972 %
{\sovast } {\bf 16}, 273.
\bibitem[\protect\citeauthoryear{Turkmani \etal}{2006}]{ture06} Turkmani, R., Cargill, P.J., Galsgaard, K., Vlahos, L., %
Isliker, H.: 2006, {\aap } {\bf 449}, 749.
\bibitem[\protect\citeauthoryear{Vlahos, Isliker, and Lepreti}{2004}]{vlae04} Vlahos, L., Isliker, H., Lepreti, F.: %
2004, {\apj }, {\bf 608}, 540. 
\bibitem[\protect\citeauthoryear{Yokoyama and Shibata}{1998}]{yosh98} Yokoyama, T., Shibata, K.: 1998, {\apj } {\bf 494}, L113.
\bibitem[\protect\citeauthoryear{Zharkova and Gordovskyy}{2004}]{zhgo04} Zharkova, V.V., Gordovskyy, M.: 2004, %
{\apj } {\bf 604}, 884.
\end{thebibliography}
\end{document}